\documentclass[amsmath,amssymb,showpacs,draft,preprint,floatfix]{revtex4}
\usepackage{graphicx}
\usepackage{latexsym}
\usepackage{amsmath}
\usepackage{amsfonts}
\usepackage{amssymb}
\begin{document}

\title{Solution for the dispersive and dissipative atom-field Hamiltonian
under  time dependent linear amplification processes}
\author{ R. Ju\'arez-Amaro,$^1$ A. Z\'u\~niga-Segundo$^2$ and H. Moya-Cessa,$^3$  }
\affiliation{$^1$ Universidad Tecnol\'ogica de la Mixteca, Apdo.
Postal 71, 69000 Huajuapan de Le\'on, Oax., Mexico\\ $^2$
Departamento de F\'{\i}sica, Escuela Superior de F\'{\i}sica y
Matem\'aticas Edificio 9, Unidad Profesional 'Adolfo L\'opez
Mateos', 07738 M\'exico, DF, Mexico
\\$^3$INAOE, Apdo. Postal 51 y 216, 72000, Puebla, Pue., Mexico}

\begin{abstract}
 The dispersive interaction between a two-level atom and a
quantized field is studied. We consider besides a time dependent
linear amplification and dissipative processes. In order to solve
the master equation for this system, we use superoperator
techniques.
\end{abstract}
\maketitle
\section{Introduction}
Recently, the geometric phase due to the Stark shift in a system
composed of a  field, driven by time-dependent linear
amplification, interacting dispersively with a two-level
(fermionic) system was studied \cite{Duzzi}. The solution for the
Hamiltonian that takes into account the above conditions, due to
its time dependence, was solved by using invariant techniques of
the Lewis-Ermakov type \cite{Lewis,Guasti}. It is well known that
dissipative dynamics must be considered when  atoms interact with
fields in the dispersive regime (far off-resonance) as the
atom-field interaction constant is replaced by a much smaller {\it
dispersive} interaction constant.

In this contribution we want to study the effect to add the
interaction with the environment for this system. we will do this
by expressing the master equation related to each element of the
density matrix and making a transformation that allows a solution
to the appropriate differential equation.

\section{Dispersive interaction}
Consider the two-level atom-field interaction Hamiltonian
\begin{equation}
H_{a-f}= \omega a^{\dagger}a+\frac{\omega_0}{2}
\sigma_z+\lambda\left( a\sigma_{+}+a^{\dag}\sigma_{-}\right),
\end{equation}
where $\lambda$ is the atom-field interaction constant, $\omega_0$
is the atomic transition frequency and $\sigma_{-}$ ($\sigma_{+}$)
is the lowering (raising) operator for the atom, with
$[\sigma_{+}, \sigma_{-}]=2\sigma_{z}$ and $a$ and $a^{\dagger}$
are the annihilation and creation operators for the cavity field,
respectively. The field  frequency is $\omega$. Note that this
interaction may also be realized in the  ion-laser interaction
\cite{Moya3a,Moya3b}.

If we consider the field frequency  far away from the atomic
transition frequency, i.e. $|\omega-\omega_0|\ge \lambda$, the
atom and the field stop exchanging energy and the Hamiltonian
above can be cast into a dispersive Hamiltonian either by using
adiabatic \cite{david} or small rotation techniques \cite{kli}.
The Hamiltonian is then written as
\begin{equation}
H_{eff}=\nu a^{\dagger}a + \frac{\omega}{2}\sigma_z+\chi
a^{\dagger}a\sigma_z+\chi \sigma_{ee},
\end{equation}
with $\sigma_{ee}=\frac{1}{2}(1-\sigma_{z})$. Now we consider the
effective Hamiltonian for the dispersive interaction between a
two-level atom and a quantized field under time dependent linear
amplification process \cite{Duzzi}

\begin{equation}
H=\nu a^{\dagger}a + \frac{\omega}{2}\sigma_z+\chi
a^{\dagger}a\sigma_z+\chi \sigma_{ee} + f(t)a^{\dagger}+f^*(t)a.
\label{H}
\end{equation}
The von Neumann equation for the  density matrix ${\mathcal R}$
taking into account the environment is
\begin{equation}
\dot{\varrho}=-i[H,{\varrho}] + {\mathcal L} {\varrho},
\end{equation}
where \cite{Arevalo}
\begin{equation}
{\mathcal L} \rho = \gamma(J-L)\rho,
\end{equation}
with
\begin{equation}
L\rho= a^{\dagger}a \rho +\rho a^{\dagger}a, \qquad J\rho=2a\rho
a^{\dagger}.
\end{equation}

Duzzioni {\it et al.} \cite{Duzzi} studied the Berry phase by
solving the Schr\"odinger equation for the Hamiltonian (\ref{H})
by using Lewis-Ermakov techniques \cite{Lewis}, commonly used to
solve time dependent harmonic oscillator interactions
\cite{Guasti}. Here we show how to solve this interaction with a
different method: a simple transformation that allows solution
even when losses are taken into account.
\section{Solution to the master equation}
We can simplify the  master equation by transforming it via
${\varrho}=\exp[-it(\frac{\omega}{2}\sigma_z+\chi \sigma_{ee}+\nu
a^{\dagger}a)]\rho\exp[it(\frac{\omega}{2}\sigma_z+\chi
\sigma_{ee}+\nu a^{\dagger}a)]$ such that we obtain
\begin{equation}
H_T=\chi a^{\dagger}a\sigma_z +
f_{\nu}(t)a^{\dagger}+f_{\nu}^*(t)a,
\end{equation}
with $f_{\nu}(t)=f(t)e^{i\nu t}$.
\begin{equation}
\dot{\rho}=-i[H_T,\rho] +{\mathcal L} \rho\label{ME}
\end{equation}

Writing the master equation for each element of the density matrix
we have
\begin{equation}
\dot{\rho}_{ee}=[R+S(f_{\nu})+{\mathcal L}]\rho_{ee}, \label{MEe}
\end{equation}
\begin{equation}
\dot{\rho}_{gg}=[-R+S(f_{\nu})+{\mathcal L}]\rho_{gg}, \label{MEg}
\end{equation}
\begin{equation}
\dot{\rho}_{eg}=[S(f_{\nu})+{\mathcal L}-i\chi L]\rho_{eg},
\label{MEeg}
\end{equation}
and
\begin{equation}
\dot{\rho}_{ge}=[S(f_{\nu})+{\mathcal L}+i\chi L]\rho_{ge},
\label{MEge}
\end{equation}
 with
\begin{equation}
R\rho=-i\chi a^{\dagger}a \rho +i\rho\chi a^{\dagger}a,
\end{equation}
and
\begin{equation}
S(f_{\nu})\rho=-i[f_{\nu}(t)a^{\dagger}+f_{\nu}^*(t)a]\rho+
i\rho[f_{\nu}(t)a^{\dagger}+f_{\nu}^*(t)a].
\end{equation}
Note that the relevant commutators are
\begin{equation}
[S(\epsilon),{\mathcal L}]\rho=S(\epsilon)\rho,
\end{equation}
\begin{equation}
[J, L]\rho=2J\rho,
\end{equation}
and
\begin{equation}
 [R,J]\rho=[R,L]\rho =0 .
\end{equation}

\subsection{Solution for $\rho_{ee}$}
 We first transform (\ref{MEe}) with
$\rho_{ee}=\exp\{(R+{\mathcal L})t\}\tilde{\rho}_{ee}$ to obtain
\begin{equation}
\dot{\tilde{\rho}}_{ee}=e^{\gamma
t}S(f_{\nu+\chi})\tilde{\rho}_{ee}=-i\left[[g_+(t)a^{\dagger}+g_+^*(t)a],\tilde{\rho}_{ee}\right]
\end{equation}
with $g_+(t)=f(t)e^{i(\nu+\chi)t+\gamma t}$ with solution
\begin{equation}
\tilde{\rho}_{ee}(t)=D^{\dagger}[iG_+(t)]
\tilde{\rho}_{ee}(0)D[iG_+(t)]
\end{equation}
with $G_+(t)=\int_0^t g_+(t) dt$ and $D(\beta)=e^{\beta
a^{\dagger}-\beta^* a}$ the Glauber displacement operator
\cite{Glauber}.
\subsection{Solution for $\rho_{gg}$}
We follow the solution for $\rho_{ee}$ above and transform
(\ref{MEg}) with $\rho_{gg}=\exp\{(-R+{\mathcal
L})t\}\tilde{\rho}_{gg}$ to obtain
\begin{equation}
\dot{\tilde{\rho}}_{gg}-i\left[[g_-(t)a^{\dagger}+g_-^*(t)a],\tilde{\rho}_{gg}\right]
\end{equation}
with $g_-(t)=f(t)e^{i(\nu-\chi)t+\gamma t}$ with solution
\begin{equation}
\tilde{\rho}_{gg}(t)=D^{\dagger}[iG_-(t)]
\tilde{\rho}_{gg}(0)D[iG_-(t)]
\end{equation}
with $G_-(t)=\int_0^t g_-(t) dt$.

\subsection{Solution for $\rho_{eg}$ and $\rho_{ge}$}
The solution for $\rho_{eg}$ (or $\rho_{ge}$) is more complicated
because it involves non Hermitian operators. Several
straightforward (but tedious) transformation to simplify the
equation may be performed. We start with
$\rho_{eg}=\exp\left(-\frac{\gamma}{2\beta}J\right)
\rho_{eg}^{(1)}$ with $\beta=\gamma + i\chi$ to obtain
\begin{equation}
\dot{\rho}_{eg}^{(1)}=[S(f_{\nu})+\frac{\gamma}{2\beta}S_1-\beta
L]\rho_{eg}^{(1)}
\end{equation}
with
\begin{equation}
S_1\rho=-2i(f_{\nu}\rho a^{\dagger}- f_{\nu}^*a\rho).
\end{equation}
By applying
\begin{equation}
\rho_{eg}^{(2)}=e^{\beta t a^{\dagger}a} \rho_{eg}^{(1)}e^{\beta t
a^{\dagger}a}
\end{equation}
we end up with the equation
\begin{equation}
\dot{\rho}_{eg}^{(2)}=-i[F_1(\beta)a^{\dagger} +F_2(\beta)a ]
+i\rho_{eg}^{(2)}[F_3(\beta)a^{\dagger}+F_4(\beta)a]
\end{equation}
with $F_1(\beta)=f_{\nu}e^{\beta t}$,
$F_2(\beta)=f_{\nu}^*e^{-\beta t}(1-\gamma/\beta)$,
$F_3(\beta)=f_{\nu}e^{-\beta t}(1-\gamma/\beta)$ and
$F_4(\beta)=f_{\nu}^*e^{\beta t}$. This equation looks now easy to
integrate. With $G_j(\beta,t)=\int F_j(\beta) dt$ we write the
solution to the above equation as
\begin{eqnarray}
& &
\rho_{eg}^{(2)}(t)=e^{-\int_0^t\left(F_2(\beta)G_1(\beta)+F_3(\beta)G_4(\beta)\right)dt}\times
\nonumber \\
& &
e^{-iG_1(\beta)a^{\dagger}}e^{-i[G_2(\beta)-G_2(\beta;0)]a}e^{iG_1(\beta;0)a^{\dagger}}
\rho_{eg}^{(2)}(0)\nonumber \\
& &
e^{-iG_4(\beta;0)a}e^{i[G_3(\beta)-G_3(\beta;0)]a^{\dagger}}e^{iG_4(\beta)a}
\end{eqnarray}
and
\begin{eqnarray}
& & \rho_{eg}^{(1)}(t)=
e^{-\int_0^t\left(F_2(\beta)G_1(\beta)+F_3(\beta)G_4(\beta)\right)dt}\times
\nonumber \\ & & e^{-\beta ta^{\dagger}a}
e^{-iG_1(\beta)a^{\dagger}}e^{-i[G_2(\beta)-G_2(\beta;0)]a}e^{iG_1(\beta;0)a^{\dagger}}
\rho_{eg}^{(1)}(0)\nonumber \\ & &
e^{-iG_4(\beta;0)a}e^{i[G_3(\beta)-G_3(\beta;0)]a^{\dagger}}e^{iG_4(\beta)a}e^{-\beta
ta^{\dagger}a}
\end{eqnarray}

The solution for $\rho_{ge}$ is similar to the one for $\rho_{eg}$
but taking $\beta \rightarrow \beta^*$.
\section{Conclusions}

We have studied the dispersive interaction between a two-level
atom and an electromagnetic field in the presence of dissipation
and time dependent linear amplification processes. By transforming
the master equation \cite{lasphys} we have managed to produce
simpler master equations for each element of the density matrix,
which we have shown to be solvable. Systems like the ones studied
here are of interest in the reconstruction of quasiprobability
distribution functions to measure the quantum state of light
\cite{Roversi1,Roversi2}.


\begin{thebibliography}{88}
\bibitem{Duzzi} E.I. Duzzioni, C.J. Villas-Boas, S.S. Mizrahi,
M.H.Y. Moussa and R.M. Serra, Europhys. Lett. {\bf 72}, 21 (2005).
\bibitem{Lewis} H.R. Lewis, Phys. Rev. Lett. \textbf{18}
510 (1967)
\bibitem{Guasti} M. Fern\'andez Guasti and H. Moya-Cessa, J. of Phys.  A {\bf 36}, 2069
(2003); H. Moya-Cessa and M. Fern\'andez Guasti, Phys. Lett. A
{\bf 273}, 1 (2003).
\bibitem{Moya3a}  H. Moya-Cessa and P. Tombesi,  Phys. Rev. A {\bf 61}, 025401 (2000).
\bibitem{Moya3b} H. Moya-Cessa, D. Jonathan and P.L. Knight,{   J. of Mod. Optics} {\bf  50}, 265 (2003).
\bibitem{david} L.Davidovich, in New Perspectives on Quantum Mechanics, Latin-American School of
Physics XXXI ELAF, (AIP Conference Proceedings, 1998) ed. by S.
Hacyan, R. J\'auregui and R. L\'opez-Pe\~na.
\bibitem{kli} A. Klimov and L.L. S\'anchez-Soto, Phys. Rev. A, {\bf 61}, 063802 (2000).
\bibitem{Arevalo} L.M. Ar\'evalo-Aguilar and H. Moya-Cessa,  Quantum. and
Semiclass. Opt. 10, 671 (1998); Rev. Mex. Fis. {\bf 42}, 675
(1996).
\bibitem{Glauber} R.J. Glauber,  {\it  Phys. Rev.}  {\bf 131}
2766, (1963).
\bibitem{lasphys} R. Ju\'arez-Amaro, J.M. Vargas-Mart\'{\i}nez and H. Moya-Cessa, Laser Physics {\bf 18},
344(2008).
\bibitem{Roversi1} H. Moya-Cessa, S.M. Dutra, J.A. Roversi, and A.
Vidiella-Barranco, { J. of Mod. Optics} {\bf 46}, 555 (1999).
\bibitem{Roversi2} H. Moya-Cessa, J.A. Roversi, S.M. Dutra, and A. Vidiella-Barranco,
Phys. Rev. A {\bf 60},  4029 (1999).
\end{thebibliography}
\end{document}